# Approaching intrinsic threshold breakdown voltage and ultra-high gain in graphite/InSe Schottky photodetector

*Zhiyi Zhang[1†], Bin Cheng[2†], Jeremy Lim[3†], Anyuan Gao[1†], Lingyuan Lyu[1, 4], Tianju Cao[1], Shuang Wang[1], Zhu-An Li[1], Qingyun Wu[3], L. K. Ang[3], Yee Sin Ang[3]\*, Shi-Jun Liang[1]\*, Feng Miao[1]\**

Z. Zhang, A. Gao, L. Lyu, T. Cao, S. Wang, Z. Li, Prof. S. J. Liang, Prof. F. Miao
National Laboratory of Solid State Microstructures, School of Physics, Collaborative Innovation Center of Advanced Microstructures, Nanjing University, Nanjing 210093, China

Prof. B. Cheng
Institute of Interdisciplinary Physical Sciences, School of Science, Nanjing University of Science and Technology, Nanjing 210094, China

J. Lim, Q. Wu, Prof. L. K. Ang, Prof. Y. Ang
Science, Mathematics and Technology, Singapore University of Technology and Design, 8 Somapah Road, Singapore 487372

L. Lyu
Physics Department, Harvey Mudd College, Claremont, CA 91711, USA

†These authors contributed equally to this work: Zhiyi Zhang, Bin Cheng, Jeremy Lim, Anyuan Gao
\* Correspondence Email:
yeesin_ang@sutd.edu.sg; sjliang@nju.edu.cn; miao@nju.edu.cn

**Keywords:** InSe, Van der Waals (vdW) Schottky junction, Avalanche photodiode, layered material

**Abstract:**

Realizing both ultra-low breakdown voltage and ultra-high gain has been one of the major challenges in the development of high-performance avalanche photodetector. Here, we report that an ultra-high avalanche gain of $3\times10^5$ can be realized in the graphite/InSe Schottky photodetector at a breakdown voltage down to 5.5 V. Remarkably, the threshold breakdown voltage can be further reduced down to 1.8 V by raising the operating temperature, approaching the theoretical limit of $1.5\mathcal{E}_g/e$ with $\mathcal{E}_g$ the band gap of semiconductor. We develop a two-dimensional impact ionization model and uncover that observation of high gain at low breakdown voltage arises from reduced dimensionality of electron-phonon (e-ph) scattering in the layered InSe flake. Our findings open up a promising avenue for developing novel weak-light detectors with low energy consumption and high sensitivity.

Avalanche photodetectors (APDs) enable detection of ultra-low light levels via charge amplification mechanism and have been widely used in many fields including weak light detection,[1-3] quantum computing[4-6] and communication.[7,8] In conventional APDs, the charge amplification mechanism is based on one-carrier cascade impact ionization process involving only one type of carriers. To achieve pronounced charge amplification, *i.e.* high avalanche gain, a large breakdown voltage is required to provide enough energy for each injected carrier to produce multiple cascade ionizations in an avalanche region defined by a length of multiple mean-free paths. This leads to a grand challenge that the ultra-high avalanche gain and the low breakdown voltage cannot be realized simultaneously in the conventional APD materials. Moreover, the breakdown voltages reported to date in the experimental works have never approached the theoretical limit of breakdown voltage $1.5\mathcal{E}_g/e$ with high gain, hindering the development of APDs with both low energy consumption and high sensitivity. Searching for novel APD materials with alternative mechanisms to realize charge amplification represents a highly promising solution for addressing such challenges.

Recently, the emerging family of two-dimensional materials and van der Waals (vdW) heterostructures has prompted a revolution in developing high-performance avalanche photodetector due to their unique properties.[3,11-13] In particular, the enhanced Coulomb interaction resulting from the quantum confinement in vdW layered materials could boost the ionization rate[12,14] during the process of impact ionization. Here, we propose a new type of APDs based on vdW Schottky junction, and realize both intrinsic threshold breakdown voltage of $1.5\mathcal{E}_g/e$ and ultra-high avalanche gain up to ~ $3\times10^5$. Such an excellent performance of the vdW Schottky APD can be well explained by a two-dimensional avalanche model. In addition, we find the temperature dependence of the breakdown voltage and the gain relies not only on the ionization process, but also on the thermally assisted carrier collection process. Our work highlights the potential of vdW Schottky junction for developing next-generation high-performance APDs.

As schematically shown in Figure 1(a), vdW Schottky APD was fabricated based on graphite/InSe vdW heterostructures, which were assembled by dry transfer technique. Here, InSe has an ultra-high in-plane mobility (~$10^4$ cm$^2$V$^{-1}$s$^{-1}$ at 4 K and ~4000 cm$^2$V$^{-1}$s$^{-1}$ at room temperature),[15] which ensures a rapid acceleration of carriers to facilitate the impact ionization process. In addition, multilayer InSe possesses a direct band gap of approximately 1.2 eV and is promising to achieve a high quantum efficiency.[11,16] Meanwhile, by employing graphite flake as the van der Waals Schottky contact, we can achieve atomically flat interface and thus avoid the interface quality degradation which is inevitable in conventional metal contact. Such high interface quality can significantly suppress the dark current and thus is critical to achieve high gain. [7,8,17] In addition, due to the outstanding conductivity of graphite, the bias voltage drops mainly on the InSe for ionization, thus enhancing the electric field in the Schottky junction and realizing avalanche breakdown at a small bias voltage. We first mechanically exfoliated InSe flakes (30~50 nm) onto highly p-doped silicon substrates covered by 300nm SiO$_2$, and graphite flakes (5~20 nm) onto polydimethylsiloxane (PDMS) sheet. Then the graphite flakes were transferred on top of InSe flakes to construct graphite/InSe Schottky junction. Standard e-beam lithography followed by e-beam evaporation was employed to fabricate 5 nm Ti/45 nm Au electrodes. The gate voltage $V_{gs}$ was applied to deplete the free charge carriers, thus providing a wide depletion region to facilitate the avalanche breakdown (see Figure S1, Supporting Information). Note that the electrical transport measurements were performed with the graphite/InSe Schottky junction reverse-biased and InSe/Ti Schottky junction forward-biased, unless specified otherwise.

We first studied the current variation ($I_{ds}$) and multiplication as a function of drain-source voltage ($V_{ds}$) in a typical device, with the results shown in Figure 1(b). The current measured at the dark condition is on the order of picoampere at low $V_{ds}$ and is at least six orders of magnitude smaller than the dark current in conventional APDs,[7,8] which demonstrates the benefit of vdW Schottky contact. Remarkably, the measured current exhibits an abrupt increase of more than five orders above a certain threshold voltage, i.e. $V_{bd}$ = 5.5 V, suggesting that an avalanche breakdown occurs. It should be noted that the graphite/InSe APD can exhibit avalanche breakdown even at room temperature (see Figure S2, Supporting Information). By illuminating the InSe flake by a laser with a wavelength of 532 nm at 6.9 pW, we observed a sharp rise in the photocurrent at a smaller threshold voltage, i.e. $V_{ds}$ = 5.1 V. The reduction in $V_{ds}$ is due to the fact that photo-injected carriers have larger initial kinetic energy than that at dark condition for the impact ionization. Based on the McIntyre's formula,[18] the multiplication factor $M$ ($M = (I_{ph} - I_{dark})/I_{ug}$ with $I_{ug}$ the photocurrent at $M$ = 1), i.e., avalanche gain, is obtained (Figure 1(b)), reaches up to $3\times10^5$. This high value at laser power of 6.9 pW corresponds to a highly-sensitive detection of 360 photons (see Section 3, Supporting Information). Notably, the light responsivity and quantum efficiency of our device were measured to be $1.16 \times 10^5 $A/W and $2.7 \times 10^5$, respectively, indicating that graphite/InSe Schottky photodetector exhibits a high sensitivity and thus facilitates the detection of weak light signals. Compared to traditional APDs,[7,8,19] our devices exhibit three orders of magnitude higher avalanche gain, and five orders of magnitude lower dark current at the same breakdown voltage. These advantages highlight the great potential of our InSe/graphite APDs in practical applications requiring low energy consumption and high sensitivity. As the laser power ($P_{opt}$) is raised (Figure 1(c) and 1(d)), the measured photocurrent as a function of reverse bias voltage exhibits two distinct features. Within the regime of low $V_{ds}$, the photocurrent gradually increases with the light intensity. Such increase of photocurrent can be attributed to the photoconductive effect that the electrical conductivity is enhanced by the photo-generated free electrons and holes in InSe. Meanwhile, in the regime that $V_{ds}$ is above the breakdown voltage and the laser power is low, i.e. below

6.9 µW, the photocurrent is almost unchanged as we increase the light intensity, as shown in Figure 1(c) and (d). We attribute the unconventional *I-V* characteristics under different light illumination to the limitation of series resistance.[20-22]

The experimental result of such high gain up to $10^5$ at a low breakdown voltage cannot be explained by the traditional model of avalanche breakdown. In the conventional APDs, the charge multiplication mechanism is based on one-carrier cascade ionization process, and therefore, an excessively high breakdown voltage of at least 30 V is required for achieving a gain of $10^5$ (see Section 4, Supporting Information), while it can be accessed at $V_{bd}$ = 5.1 V in our observations. We also carefully checked the channel length dependence of the breakdown voltage in the InSe-based APDs, showing similar breakdown voltages for different channel lengths (see Figure S4, Supporting Information), which is also inconsistent with the model of one-carrier impact ionization.[23]

We attribute our observation to the quantum confinement effect induced by the vdW gap in the layered InSe, which facilitates an alternative mechanism, *i.e.* two-carrier process.[24,25] As shown in Figure 2(a), the vdW gap acts as a tunneling barrier of ~1.85 eV (see Figure S5, Supporting Information) and suppresses the out-of-plane charge transport,[26,27] leading to the reduction in the dimensionality of e-ph scattering and the enhancement of the Coulomb interaction.[12,14] In this sense, the energy dissipation *via* e-ph scattering will be suppressed, and the impact ionization rate will be enhanced,[14] enabling the two-carrier impact ionization process which can explain the observation of a much higher gain at a much lower breakdown voltage compared to previous works (see Figure S22, Supporting Information). Specifically, as schematically shown in Figure 2(b), when an electron is injected into vdW Schottky junction and accelerated under the action of both external and built-in electrical field to initialize an impact ionization event, an electron-hole pair will be generated. Then the generated hole will drift back to produce another electron-hole pair and the two electrons are collected by the electrode. As the cycle repeats, more and more carriers are generated in a region within the length of only one mean-free path $\lambda$ and collected through electrodes, which explains the channel length independence of breakdown voltage in our device. Assuming electrons and holes have equal ionization probability *p*, the avalanche gain for two-carrier impact ionization process can be determined by *M*

= $p/(1-p)$.[24] Accordingly, if $p$ approaches 1 due to the reduction in the dimensionality of e-ph scattering, ultra-high avalanche gain can be reached at an extremely low breakdown voltage, which is consistent with our results.

To further investigate the ionization process in layered InSe, we carried out theoretical calculations considering the effect of reduced dimensionality of e-ph scattering. Note that the ionization probability $p$ is usually expressed by the ionization rate per unit path length $\alpha$ (*i.e.*, $p = \alpha \times \lambda$),[9] which is an intrinsic property of semiconducting material characterizing the multiplication process.[28] Assuming that the carrier transport in InSe are strongly confined within the 2D atomic layers while the carrier transfer between adjacent layers can be neglected, we find that the two-dimensional ionization rate $\alpha_{2D}$ can be described by

$$\alpha_{2D} \propto \frac{eE}{\mathcal{E}_{K,th}}. \tag{1}$$

Here $E$ is the electric field strength, and $\mathcal{E}_{K,th}$ is the threshold carrier kinetic energy ($\mathcal{E}_{K,th} = 1.5\mathcal{E}_g$) above which the charge multiplication occurs. Here we adopted the methods developed by Wannier[29] and Wolff[28] for solving the Boltzmann equation in the high-field limit, and generalized these methods to a two-dimensional geometry (see Section 8, Supporting Information). Assuming there is negligible energy loss to phonons per collision, we only need to solve for the zeroth and first-order terms of the carrier velocity distributions, which in turn allow us to derive an expression for the two-dimensional ionization rate $\alpha_{2D}$. In the limit where the kinetic energy gained by the carrier accelerated over $\lambda$ by the field $E$ far exceeds the energy lost to phonons (i.e., $\hbar\omega \ll eE\lambda$), we arrive at Equation 1 above.

Based on this developed model, we theoretically examine mean-free path $\lambda$ and electrical field $E$ dependence of the ionization rate $\alpha_{2D}$ and gain $M$, with results shown in Figure 2(c) and 2(d), respectively. The gain $M$ shown in Figure 2(d) is given by multiplying the gain obtained from a single injected electron by the layer number (~40) of InSe flake used in our experiment. Note that there exists a white region in Figure 2(d) where $M$ diverges, which corresponds to the regime where the ionization probability $p \geq 1$. While seeming unphysical, this simply implies that carriers are sufficiently energetic for impact ionization to occur with certainty. Although the divergent *M* cannot be accessed in our experiment due to the limitation of resistance, it is in good agreement with the ultra-high gain observed in our experiments.

To further demonstrate that the reduced dimensionality of e-ph scattering indeed plays a significant role in the ionization rate, we compare $\alpha_{2D}$ with the ionization rate $\alpha_{3D}$ for bulk geometries whose carrier motion is isotropic in three dimensions. By fixing $\lambda = 22$ nm which is the value extracted from our experiments (see Section 7, Supporting Information), we obtain $\alpha_d$ (where $d$ denotes either 2D or 3D) and the gain $M$ obtained from a single injected electron as functions of *E*, as shown in Figure 2(e) and 2(f). We find that $\alpha_{2D}$ is asymptotically described by Eq. (1) and reaches $p = 1$ (black dotted line) at high fields, as shown in Figure 2(e). In contrast, at large *E*, $\alpha_{3D}$ saturates to $\frac{C}{\lambda}$ with C = $\sqrt{3}/2$ (see Equation S73, Supporting information), indicating a $p = \lambda \alpha_{3D}$ smaller than 1 for all *E*, which inherently limits the possible increase in gain with a corresponding increase in *E* (Figure 2(f)). The divergent gain for the two-dimensional case and the saturation for the three-dimensional case in turn support the foregoing hypothesis that the ultra-high gain observed in our vdW device results from the reduced dimensionality of the e-ph scattering in the layered material.

According to our model, the suppressed energy loss to lattice and the increased ionization rate in the layered InSe allow for the realization of intrinsic threshold breakdown voltage approaching the theoretical limit of $1.5\mathcal{E}_g/e$ accompanied with a high gain. To search for this intrinsic threshold breakdown voltage, we varied the temperature to continuously tune the lattice vibration, which can modify e-ph scattering strength and thus the breakdown voltage.[30] Figure 3(a) plots the drain-source current $I_{ds}$ as a function of bias voltage $V_{ds}$ at different temperatures ranging from 140 K to 260 K. Remarkably, we observed a breakdown voltage as low as 1.8 V at T = 260 K, which approaches the theoretical limit of threshold breakdown voltage $1.5\mathcal{E}_g/e$ by using the bandgap (i.e. $\mathcal{E}_g = 1.2$ eV) of InSe. This is the first time that an intrinsic threshold breakdown voltage has been observed experimentally. We attribute this to the outstanding conductivity of the graphite contact, which makes the bias voltage dropping mainly on InSe and thus facilitate the ionization. At this breakdown voltage, we achieved an avalanche gain around 110. It is worth noting that a breakdown voltage of at least 12 V is required for conventional materials to access such a high gain (see Section 4, Supporting Information), indicating that the use of vdW material gives an unprecedented opportunity for addressing the traditional challenge that the intrinsic threshold breakdown voltage and the high avalanche gain cannot be simultaneously achieved.

We now turn to the temperature dependence of the breakdown voltage and the avalanche gain. As shown in Figure 3(b), the breakdown voltage monotonically decreases with the rising temperature while the gain exhibits non-monotonic temperature dependence, both of which are fundamentally different from the features widely reported in the conventional APDs. To interpret such unusual temperature dependence of the breakdown voltage and the gain, we investigate three consecutive electron transport processes, *i.e.*, I. injection, II. ionization, and III. collection,[31] as schematically shown in the inset of Figure 3(a). First, at high temperature, the ionization process is suppressed since the mean-free path is reduced due to enhanced lattice vibrations. In this case, a larger breakdown voltage is required to provide stronger electrical field to accelerate electrons to obtain the threshold energy, which is inconsistent with the negative temperature coefficient of $V_{bd}$ shown in Figure 3(b). In contrast, in the injection and collection processes, as electrons possess more thermal energy with the increasing temperature, the electric field strength required to pass over the potential barrier decreases. Moreover, the number of injected electrons in the injection process are much less than that in the collection process, indicating a negligible temperature effect in the injection process. Thereby, we can attribute the negative temperature coefficient of breakdown voltage to the temperature-enhanced collection process through InSe/Ti Schottky junction. Different from the collection-dictated monotonical temperature dependence of the breakdown voltage, the non-monotonical temperature coefficient of the gain is governed by the competition between ionization and collection. As the temperature rises from 80 K to 180 K, the gain is limited by the collection efficiency, which is consistent with the experimental result shown in Figure 3(a) that the reverse saturation current increases as temperature increases. As the temperature is further raised, *i.e.* from 180 K to 260 K, the mean-free path is significantly reduced and the ionization process becomes the major limiting factor, despite the collection efficiency continues to increase.

The underlying physics described above can also be justified by reverse-biasing the InSe/Ti Schottky junction, which leads to a positive temperature dependence of the breakdown voltage ranging from 80 K to 180 K and a decrease in gain from the lower temperature of 100 K, as shown in Figure 3(c) and 3(d). Note that the InSe/graphite

Schottky barrier height is much smaller than that in the InSe/Ti junction, as supported by the thermionic emission measurement shown in Figure S16 (Supporting information), making the ionization generated electrons pass through the InSe/graphite barrier, i.e., the collector terminal, much easier. In this case, the impact ionization would be the major factor for determining the breakdown voltage and the gain, and thus leads to a positive temperature dependence of breakdown voltage and a decrease in gain from a lower temperature, as we observed. To further systematically investigate the temperature coefficients for different Schottky barrier height, we have designed three sets of control experiments, with the result confirming that the sign of temperature coefficient is a result of the competition between the collection and ionization processes (see Figure S15, Supporting Information).

In summary, we observed an intrinsic threshold breakdown voltage of 1.8 V as well as ultra-high avalanche gain ~ $3 \times 10^5$ in the graphite/ InSe Schottky APD. Based on a two-dimensional carrier multiplication model, we attribute the results to the emergence of two-carrier ionization process that is facilitated by the reduced dimensionality of e-ph scattering in layered InSe semiconductor. In addition, we demonstrate that both the breakdown voltage and avalanche gain not only depend on the temperature dependent ionization, but are also affected by the thermally assisted collection process. These findings reveal the fundamental difference of carrier multiplication physics between conventional and layered semiconductors, and open up new perspectives for future APDs with both low energy consumption and high sensitivity.

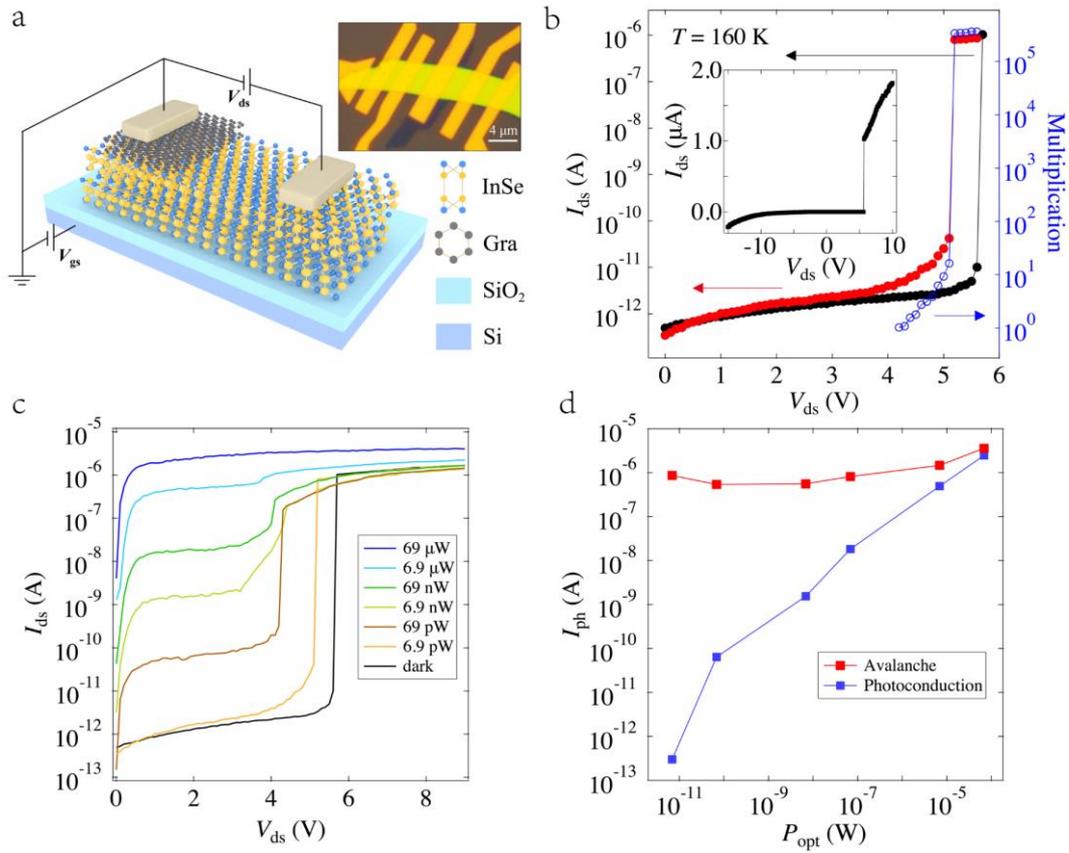

**Figure 1.** (a) Schematic of the InSe/graphite APD. Inset: optical image of the graphite/InSe APD. (b) Logarithmic scale $I_{ds}$-$V_{ds}$ characteristics (black line, dark; red line, 532 nm laser illuminated with 6.9 pW), and the corresponding multiplication factor (blue line) at 160 K. The corresponding axes are denoted by the arrows. Inset: linear scale $I_{ds}$-$V_{ds}$ characteristic covering negative $V_{ds}$ in dark condition. (c) $I_{ds}$-$V_{ds}$ curves measured at a range of laser powers from 6.9 pW to 69 μW. (d) Photocurrent $I_{ph}$ of avalanche mode at $V_{ds}$ = 5.5 V (red line) and photoconductive mode at $V_{ds}$ = 2 V (blue line) as functions of laser power.

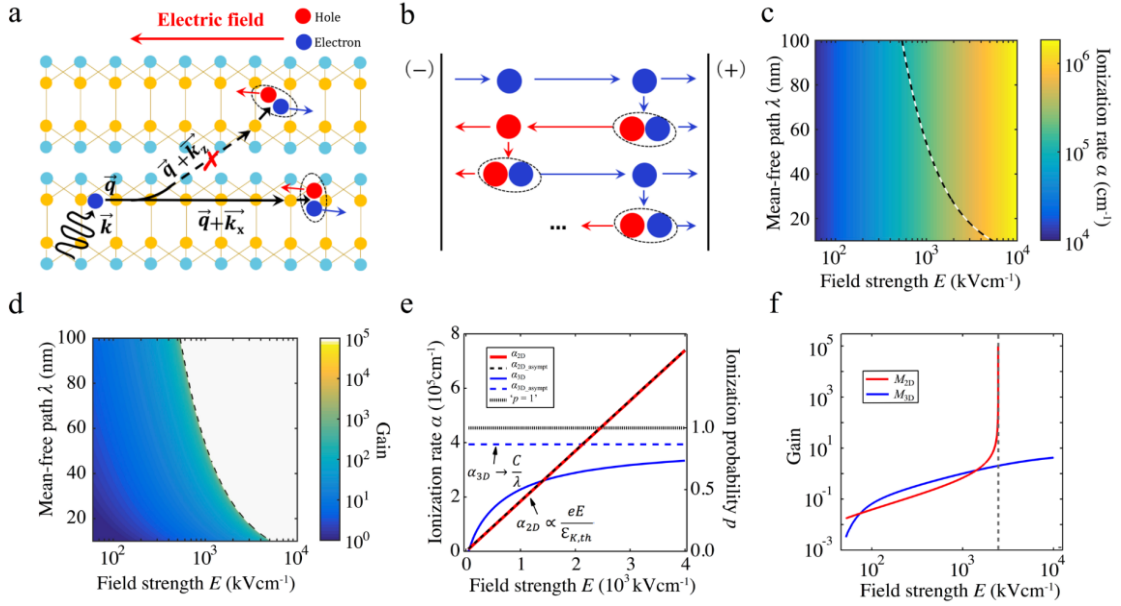

**Figure 2.** (a) Impact of reduced dimensionality of e-ph scattering on the electron acceleration process. (b) Illustration of the two-carrier impact ionization process. (c) Ionization rate $\alpha_{2D}$ and (d) Gain as functions of electric field strength $E$ and mean-free path length $\lambda$. The black dashed line indicates that impact ionization probability ($p = \alpha \times \lambda$) is equal to 1 and gain $M$ is divergent. (e) $\alpha$ and $p$ as functions of field strength $E$ in two dimensions (red solid line) and three dimensions (blue solid line). The black dashed lines and blue dashed lines are the asymptotes of $\alpha_{2D}$ and $\alpha_{3D}$, respectively. The horizontal black dotted line corresponds to the case where $p = 1$. (f) Gain as functions of $E$ in two dimensions (red solid line) and three dimensions (blue solid line). The vertical grey dashed line corresponds to the case where gain $M$ is divergent.

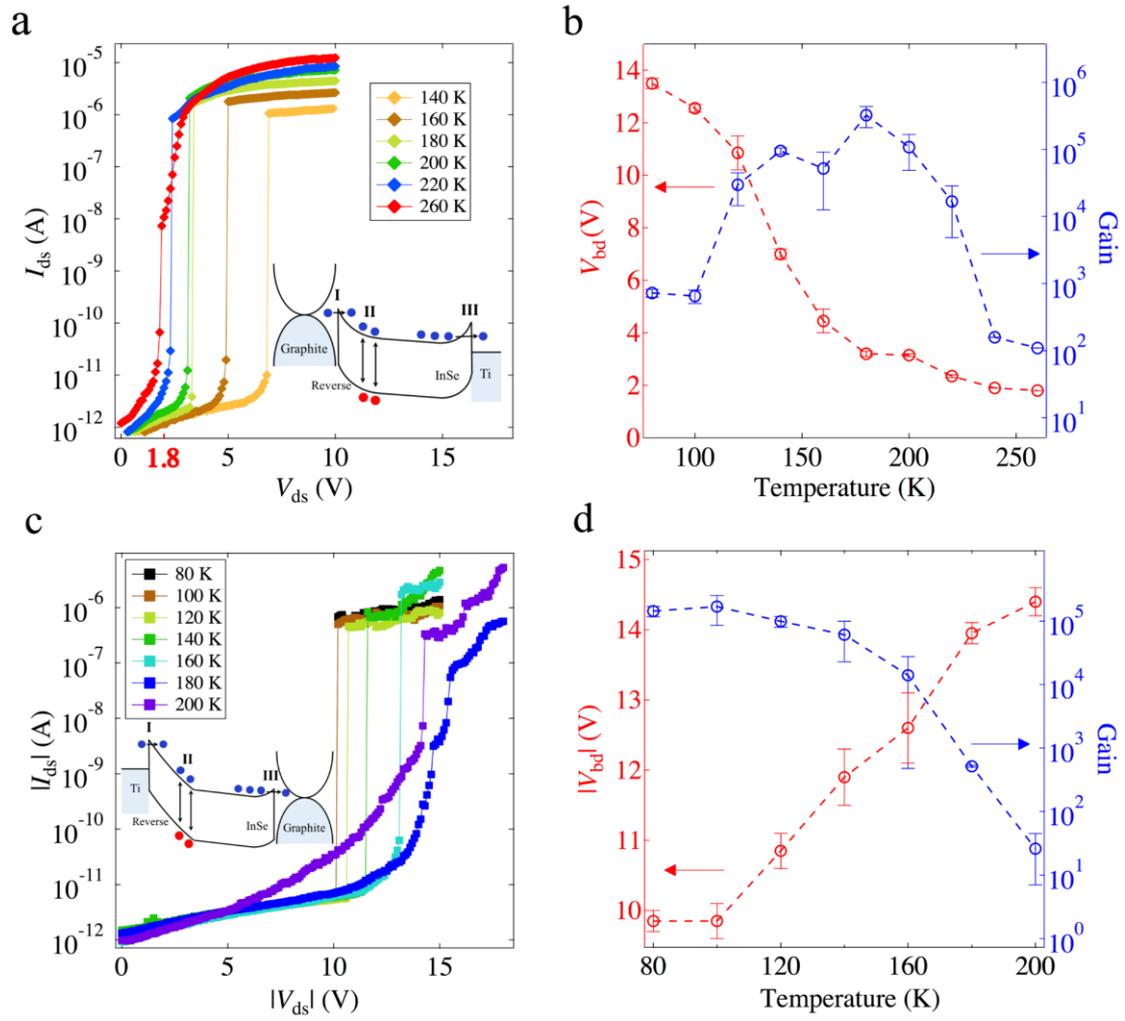

**Figure 3.** (a) $I_{ds}$-$V_{ds}$ curves at different temperatures. Inset: three electron transport processes (I. injection, II. ionization, and III. collection). (b) Breakdown voltage ($V_{bd}$) and gain as functions of temperature. (c) $I_{ds}$-$V_{ds}$ curves and (d) Breakdown voltage ($V_{bd}$) and gain as functions of temperature with the InSe/Ti Schottky junction reverse-biased.

**Experimental Section**

*Device fabrication and measurement setup.* We first mechanically exfoliated InSe flakes (30~50 nm) onto highly p-doped silicon substrates covered by 300nm $SiO_2$, and graphite flakes (5~20 nm) onto polydimethylsiloxane (PDMS) sheet. Then the graphite flakes were transferred on top of InSe flakes to construct graphite/InSe Schottky junction. Standard e-beam lithography followed by e-beam evaporation was employed to fabricate 5 nm Ti/45 nm Au electrodes. The $I_{ds}$-$V_{ds}$ curve of the Ti/InSe/graphite were measured when the gate voltage $V_{gs}$ = -40 V (Figure 1 and Figure 3). All the electrical measurements were performed by using a Keithley 2636B dual-channel digital source meter. The SAPPHIRE 532-20 CW SF CDRH laser was used as the light source, the spot size of which was focused to 1 $\mu m$ in diameter. The change of temperature in the electric measurements were performed with an Oxford instrument Microstate HiRes.

*Characterizations.* The thickness of the sample was determined by the atomic force microscope (AFM) using Bruker Multimode 8. The Raman spectra of the sample were acquired by using WiTec alpha300 under 532 nm excitation. The transfer curve of InSe and the $I_{ds}$-$V_{ds}$ curve of the Ti/InSe/graphite were acquired using a Keithley 2636B source meter.

*Calculations of ionization rate for 2D material.* We first adopted the methods developed by Wannier [29] and Wolff [28] for solving the Boltzmann equation in the high-field limit, and generalized these methods to a two-dimensional (2D) geometry. Then we used the approach undertaken by Wolff to solve for the velocity distributions in 2D case. Finally, by solving for the velocity distribution, we can obtain expressions for the 2D ionization rate per unit length as well as the gain $M$ (Figure 2(c)-(f)) (see Section 8 and 9, Supporting Information).

*Statistical Analysis.* The data is used without any transformation, except for the following: Figure 1(b), Figure 3(b) and Figure 3(d). Figure 1(b) presents the multiplication factor $M$ based on the McIntyre's formula ($M = (I_{ph}- I_{dark})/I_{ug}$ with $I_{ug}$ the photocurrent at $M = 1$). [18] In Figure 3, $I_{ds}$-$V_{ds}$ characteristic of the same graphite/InSe/Ti device were performed for two cycles at the dark condition at different temperature, where a single cycle was presented in Figure 3(a) and (c). In Figure 3(b) and Figure 3(d), the gain and breakdown voltage were given in the form of mean $\pm$ standard deviation. The statistical analysis was carried out using IGOR Pro.

**Supporting Information**

Supporting Information is available from the Wiley Online Library or from the author.

**ACKNOWLEDGEMENTS**

This work was supported in part by the National Natural Science Foundation of China (62122036, 62034004, 61921005, 61974176, 12074176), the Strategic Priority Research Program of the Chinese Academy of Sciences (XDB44000000), the National Key R&D Program of China (2019YFB2205400 and 2019YFB2205402) and F.M. acknowledges the support from the AIQ foundation. L.K.A. acknowledge the support of A*STAR AME IRG (A2083c0057). Y.S.A. acknowledge the support of SUTD Start-up Research Grant (SRG SCI 2021 163) The computational work for this work was partially performed on resources of the National Supercomputing Centre, Singapore (http://www.nscc.sg).

**Conflict of Interest**

The authors declare no competing interests.